\documentclass[aps,superscriptaddress,nofootinbib,showpacs,eqsecnum,preprint,tightenlines]{revtex4}
\usepackage{hyperref}
\usepackage{epsfig,rotating}
\usepackage{amsmath,amssymb}
\usepackage{dsfont}
\numberwithin{equation}{section}

\newcommand{\vp}{\vec{p}}

\newcommand{\vk}{\vec{k}}

\newcommand{\ta}{\tilde{a}}
\newcommand{\tFp}{\widetilde{F}}
\newcommand{\tfu}{\widetilde{f}}
\newcommand{\td}{ {\delta}}

\newcommand{\kpm}{\frac{\vec{k}\cdot \vec{p}}{m}}

\newcommand{\be}{\begin{equation}}
\newcommand{\ee}{\end{equation}}
\newcommand{\bea}{\begin{eqnarray}}
\newcommand{\eea}{\end{eqnarray}}

\begin{document}

\title{Warm dark matter at small scales:\\ peculiar velocities and phase space density.}

\author{Daniel Boyanovsky}
\email{boyan@pitt.edu} \affiliation{Department of Physics and
Astronomy, University of Pittsburgh, Pittsburgh, PA 15260}

\date{\today}

\begin{abstract}
We study the scale and redshift dependence of the
  power spectra for density perturbations
and peculiar velocities, and the evolution of a coarse grained phase
space density
 for  (WDM) particles that decoupled
during the radiation dominated stage. The (WDM) corrections are
obtained in a perturbative expansion valid  in the range of
 redshifts at which N-body simulations set up initial conditions,  and for a wide range of scales.
 The redshift dependence is determined by the
kurtosis $\beta_2$ of the distribution function at decoupling. At
large redshift there is an enhancement of peculiar velocities for
$\beta_2 > 1$ that contributes to free streaming and leads to
further suppression of the matter power spectrum and an enhancement
of the peculiar velocity autocorrelation function at   scales
smaller than the free streaming scale. Statistical fluctuations of
peculiar velocities are also suppressed on these scales by the same
effect. In the linearized approximation, the coarse grained phase
space density features redshift dependent (WDM) corrections from
gravitational perturbations  determined by the power spectrum of
density perturbations and $\beta_2$. For $\beta_2
> 25/21$ it \emph{grows
logarithmically} with the scale factor as a consequence of the
suppression of statistical fluctuations. Two specific models for WDM
are studied in detail. The (WDM) corrections relax the bounds on the
mass of the (WDM) particle candidate.
\end{abstract}


\maketitle

\section{Introduction}\label{sec:intro}

The current paradigm of structure formation, the $\Lambda CDM$
standard cosmological model, describes large scale structure
remarkably well. However, observational evidence has been
accumulating suggesting that the cold dark matter (CDM) scenario of
galaxy formation \emph{may} have problems at small, galactic, scales.

Large scale simulations seemingly yield  an over-prediction of
satellite galaxies\cite{moore2} by almost an order of
magnitude\cite{kauff,moore,moore2,klyp,will}. Simulations within the
$\Lambda$CDM paradigm also yield a density profile in virialized
(DM) halos that increases monotonically towards the
center\cite{frenk,dubi,moore2,bullock,cusps} and features a cusp,
such as the Navarro-Frenk-White (NFW) profile\cite{frenk} or more
general
  central density profiles $\rho(r) \sim r^{-\beta}$ with
$1\leq \beta \lesssim 1.5$\cite{moore,frenk,cusps}. These density
profiles accurately describe clusters of galaxies but there is an
accumulating body of observational
evidence\cite{dalcanton1,van,swat,gilmore,salucci,battaglia,deblok,kravstov}
suggesting that the     central regions of dark matter
(DM)-dominated dwarf spheroidal satellite (dSphs) galaxies
 feature smooth cores instead of cusps as predicted by (CDM). Some observations suggest\cite{salucci2}
  that the mass distribution of spiral disk galaxies can
  be best fit by a cored Burkert-type profile\cite{salucci2}.
  This difference is known as the core-vs-cusp problem\cite{deblok,kravstov}.
  The case for core-dominated halos has been recently bolstered by the analysis
   of rotation curves from the THINGS survey\cite{things}.

Warm dark matter (WDM) particles were
invoked\cite{mooreWDM,turokWDM,avila} as possible solutions to the
discrepancies both in the over abundance of satellite galaxies and
as a mechanism to smooth out  the cusped   density profiles
predicted by (CDM) simulations into the  cored profiles that fit the
observations in   (dShps). (WDM) particles feature a range of
velocity dispersion in between the (CDM) and hot dark matter leading
to free streaming scales that smooth  out small scale features and
could be consistent with core radii of the (dSphs). If the free
streaming scale of these particles is  smaller than the scale of
galaxy clusters, their large scale structure properties are
indistinguishable from (CDM) but may affect the  \emph{small} scale
power spectrum\cite{bond}   providing an explanation of the smoother
inner profiles of (dSphs) and fewer satellites.

Furthermore recent numerical results hint to more evidence of
possible small scale discrepancies with the $\Lambda CDM$ scenario:
another over-abundance problem, the ``emptiness of voids''
\cite{tikhoklypin} and the spectrum of ``mini-voids''\cite{tikho},
both of which may be explained by a WDM candidate. Constraints from
the luminosity function of Milky Way satellites\cite{maccio} suggest
a lower limit for the mass  of a WDM particle of a few $
\mathrm{keV}$, a result consistent with
 Lyman-$\alpha$\cite{lyman,lyman2,vieldwdm}, galaxy power spectrum\cite{abakousha}  and lensing observations\cite{maccio2}.
 More recently, results from the Millenium-II simulation\cite{sawala} suggest that the $\Lambda CDM$ scenario
 \emph{overpredicts} the abundance of massive $\gtrsim 10^{10}\,M_{\odot}$ haloes, which is
corrected with a WDM candidate of $m\sim 1\,\mathrm{keV}$. A model independent analysis suggests that dark matter particles
with a mass in the $\mathrm{keV}$ range is a suitable (WDM)
candidate\cite{hectornorma,hecsal}. Recent counterarguments\cite{kapli,dalal} seem to suggest that (WDM)
cannot explain cores in (LSB) galaxies, thus the controversy continues.

In absence of conclusive evidence in favor of or against cusps or
cores, and in view of the ongoing controversy and  the body of
emerging evidence in favor of (WDM), a deeper understanding of  the
small scale clustering properties of (WDM) candidates is warranted.

\vspace{2mm}

\textbf{Motivation and goals:}

 \textit{Redshift dependence of the
power spectrum and peculiar velocities:}
 recent N-body simulations of
WDM\cite{tikho,maccio} set up initial conditions   at
$z=40$\cite{maccio} or $z=50$\cite{tikho}  with a rescaled version
of the CDM power spectrum from a fit provided in ref.\cite{vieldwdm}
that inputs a cutoff from free streaming, however, these simulations
neglected the  velocity dispersion of the WDM particles in the
initial conditions. We seek to understand both the redshift
dependence of the matter and peculiar velocity power spectrum in
this range of redshifts for a wide range of scales.

\vspace{1mm}

\textit{Phase space density:} in a seminal article Tremaine and
Gunn\cite{TG} provided bounds on the mass of the DM particle from
phase space density considerations: whereas in absence of
self-gravity the fine-grained phase space density (or distribution
function) is conserved after the DM species decouples from the
plasma, phase mixing theorems\cite{theo} assert that a coarse
grained phase space density  always \emph{diminishes} as a result of
phase mixing  (violent relaxation)\cite{theo,dehnen}. Therefore the
microscopic phase space density provides an \emph{upper bound} from
which constraints on the mass can be extracted. These arguments were
generalized in
refs.\cite{dalcanton1,hogan,coldmatter,darkmatter,boysnudm,hectornorma}
to a coarse grained phase space density obtained from   moments of
the microscopic distribution function. In
ref.\cite{coldmatter,darkmatter,boysnudm,hectornorma} this coarse
grained phase space density was combined with photometric
observations of   (dShps) to constrain the mass and the number of
relativistic degrees of freedom at decoupling.

Although the microscopic phase space density, namely the
distribution function, obeys the collisionless Boltzmann equation,
the evolution of the \emph{coarse grained} phase space density is
not directly obtained from this equation (see discussion in
ref.\cite{dehnen}). Although  the \emph{proxy} phase space density
introduced in refs.\cite{dalcanton1,hogan,coldmatter,darkmatter} is
conserved after decoupling, its evolution \emph{does not} include
self-gravity. Therefore there remains the unexplored question of
precisely what happens to the microscopic phase space density or its
\emph{proxy} introduced in
refs.\cite{dalcanton1,hogan,coldmatter,darkmatter} when
gravitational perturbations are included in the Boltzmann equation.
One aspect is clear: the perturbations of the distribution function
(microscopic phase space density) feature \emph{two} moments that
\emph{grow} under gravitational perturbations: the first moment
(density perturbations) and the second moment (velocity
perturbations) which are actually related via the continuity
equation on sub-horizon scales. In this article we study the
evolution of the coarse-grained phase space density introduced in
refs.\cite{dalcanton1,hogan,coldmatter,darkmatter} as a function of
redshift and scale for (WDM) particles in order to assess how the
original arguments are modified by gravitational perturbations,
again in the regime of redshifts at which N-body simulations set up
initial conditions.

\vspace{2mm}

\textbf{Results:}

Armed with the results recently obtained in ref.\cite{smallscale} we
obtain a perturbative expansion of the \emph{redshift corrections}
to the matter, peculiar velocity power spectra and evolution of a
coarse-grained phase space density. This expansion is valid in the
regime $z\ll z_{eq}$ for a wide range of scales and is a distinct
feature of (WDM) particles. These corrections depend on the kurtosis
$\beta_2$ of the unperturbed distribution function. Peculiar
velocities contribute to the velocity dispersion and free streaming
and lead to a \emph{suppression} of the matter power spectrum for
$\beta_2 >1$ at scales smaller than the free streaming scale at
redshifts $ z \simeq 30-50$. The peculiar velocity power spectrum is
enhanced at these scales and reshift leading to an increase of the
peculiar velocity autocorrelation function and a suppression of
statistical fluctuations. For (WDM) perturbations in the linearized
approximation, it is found that the coarse grained phase space
density introduced in
refs.\cite{dalcanton1,hogan,coldmatter,darkmatter} \emph{grows
logarithmically} with the scale factor for $\beta_2 > 25/21$. Two
specific models of (WDM) particles motivated by particle physics are
studied in detail.  Implications on the bounds for the mass of the
(WDM) particle are discussed.

\section{Preliminaries}\label{sec:prelim}

We begin by establishing some notation and conventions that are used
in the analysis. Since we focus on the region of redshift $z\gg 1$ we can safely neglect the dark energy component and we consider a radiation and matter dominated
cosmology with \be H^2=  \frac{\dot{a}^2}{a^4} = H^2_0
\left[\frac{\Omega_r}{a^4}+ \frac{\Omega_m}{a^3}\right] =
\frac{H^2_0 \Omega_m}{a^4}\left[a+a_{eq}\right] \label{H2}\ee
 where the dot stands for derivative with respect to conformal time ($\eta$), the scale factor is normalized to $a_0=1$ today,  and
\be a_{eq} = \frac{\Omega_r}{\Omega_m} \simeq \frac{1}{3229}
\,.\label{aeq}\ee Introducing \be \ta = \frac{a}{a_{eq}}\,,
\label{tildea}\ee it follows that \be \frac{d\,\ta}{d\eta} =
\left[\frac{H^2_0 \Omega_m}{a_{eq}}
\right]^{\frac{1}{2}}\,\left[1+\ta\right]^{\frac{1}{2}}\,.
\label{dotta}\ee At matter-radiation equality  we define \be k_{eq}
\equiv H_{eq}\,a_{eq} = \sqrt{2}\left[\frac{H^2_0 \Omega_m}{a_{eq}}
\right]^{\frac{1}{2}}  =
 \frac{9.8 \times 10^{-3}}{\mathrm{Mpc}} \label{keq}\ee corresponding to the comoving wavevector that enters the Hubble radius at matter-radiation equality, where we have used $\Omega_m h^2 = 0.134
 $\cite{WMAP7}.

We study the evolution of  perturbations in  the conformal Newtonian
gauge

 \bea g_{00} & = &
-a^2(\eta)\Big[1+2\psi(\vec{x},\eta) \Big] \label{g00}\\ g_{ij} & =
& a^2(\eta)\Big[1-2\phi(\vec{x},\eta)\Big] ~\delta_{ij} \,.
\label{gij}\eea The perturbed distribution function is given by \be
f(p,\vec{x},\eta) = f_0(p)+F_1(p,\vec{x},\eta) \label{f}\ee where $
f_0(p) $ is the unperturbed distribution function,    which after
decoupling obeys the collisionless Boltzmann equation in absence of
perturbations and $\vec{p},\vec{x}$ are   comoving momentum and
coordinates respectively. As discussed in
ref.\cite{coldmatter,darkmatter,boysnudm} the unperturbed
distribution function is of the form \be f_0(p) \equiv
f_0(y;\chi_1,\chi_2,\cdots) \label{fform}\ee where \be y =
\frac{p}{T_{0,d}} \label{y} \ee where $p$ is the comoving momentum
and $T_{0,d}$ is the decoupling temperature \emph{today},  \be
T_{0,d} = \Big(
\frac{2}{g_d}\Big)^{\frac{1}{3}}~T_{CMB}\,,\label{T0d}\ee with $g_d$
being the effective number of relativistic degrees of freedom at
decoupling, $T_{CMB}= 2.35 \times 10^{-4}~\mathrm{eV}$ is the
temperature of the (CMB) today, and  $\chi_i$ are dimensionless
couplings or ratios of mass scales.

We neglect stress anisotropies, in which case $\phi = \psi$ and
introduce \be \tFp(\vp,\vk,\eta) =
\frac{F_1(\vp,\vk,\eta)}{n_0}~~;~~\tfu (p) = \frac{f_0(p)}{n_0}
\label{tildes} \ee where \be n_0 = \int \frac{d^3p}{(2\pi)^3}
 \,f_0(p)\,, \label{n0}\ee is the background density of (DM) \emph{today}.
  Therefore   \be \td(\vk,\eta) = \int \frac{d^3p}{(2\pi)^3} \,\tFp(\vk,\eta)\,. \label{tildelta} \ee
   becomes $\delta \rho_{m}/\rho_m$ after the DM particle becomes non-relativistic.

 Introducing spatial Fourier transforms in terms of comoving momenta $\vec{k}$ (we keep the same
 notation for the spatial Fourier transform of perturbations), and neglecting stress anisotropies
 the  linearized Boltzmann equation
  for perturbations is given by\cite{ma,dodelson,giova,lythbook,weinbergbook,ruthbook,kodama}

 \be \dot{\tFp}(\vk,\vp\,;\eta)+ i   \frac{k\,\mu\,p}{\epsilon(p,\eta)}~\tFp(\vk,\vp\,;\eta)+
 \Big(\frac{d~\tfu(p)}{dp} \Big)\Big[p~\dot{\phi}(\vk,\eta) - ik\,\mu \, \epsilon(p,\eta)~\phi(\vk,\eta) \Big] =0
   \label{BE}\ee where dots stand for derivatives with respect to conformal time, $\mu = \widehat{\mathbf{k}}\cdot \widehat{\mathbf{p}}$, and $\epsilon(p,\eta) =
   \sqrt{p^2+m^2\,a^2(\eta)} $
  is the conformal energy of the particle of mass $m$.
  The gravitational potential is determined by Einstein's equation\cite{ma,dodelson}.

As discussed in ref.\cite{smallscale} for a (WDM) particle with a
mass in the $\sim \,\mathrm{keV}$ range,  there are three stages of
evolution: I) radiation domination and the DM particle is
relativistic, II) radiation domination and the DM particle is
non-relativistic and III) the matter dominated stage, during which
cold and warm DM particles are non-relativistic.


During stages (I) and (II) the gravitational potential is completely determined by the radiation component and the Boltzmann equation for the distribution function of the WDM particle is solved by integrating eqn. (\ref{BE}) with $\phi$ being determined by the radiation component. During stage (III) the gravitational potential is determined by the matter component and the Boltzmann equation becomes a self-consistent Vlasov-type equation.

Since the Boltzmann equation is first order in time, the solution during stages (I) and (II)  becomes the initial condition for the evolution during stage (III).


In this article we focus on the evolution of peculiar velocities and
phase space density during the matter dominated stage $10 \leq z
\lesssim z_{eq}$, corresponding to stage (III) during which dark
energy can be neglected. Typical N-body simulations setup initial
conditions which input the matter power spectrum from linear
 perturbation theory at $z \simeq 30-50$.

 In this stage the WDM is non-relativistic, hence $p/\epsilon(p,\eta)= p/m\,a(\eta)$, and the Bolzmann equation simplifies by introducing the variable
 \be
s(\eta) = \int^\eta \frac{d\eta'}{a(\eta')} \equiv
\frac{2\sqrt{2}\,u(\eta)}{k_{eq}a_{eq}} \label{sdefd}\ee where the dimensionless variable
\be
u(\eta) = \frac{1}{2}\, \ln\Bigg[
\frac{\sqrt{1+\ta(\eta)}-1}{\sqrt{1+\ta(\eta)}+1}\Bigg] ~~;~~ u_{NR} \leq u(\eta) \leq 0 \,,\label{udef}\ee  is normalized   so that $u(\infty) =0$ and
 introduced\cite{smallscale}
 \be u_{NR} =   \ln\Big[\frac{\sqrt{\ta_{NR}}}{2} \Big]~~;~~\ta_{NR}=\langle  {V}^2(t_{eq}) \rangle^\frac{1}{2}\,
 \label{uNR}\ee where
 $\ta_{NR}$ corresponds to the time when the particle becomes non-relativistic, and
$\langle  {V}^2(t_{eq}) \rangle $ is the velocity dispersion of the
DM particle at matter-radiation equality  given by\cite{smallscale}
 \be \langle {V}^2(t_{eq})
\rangle^\frac{1}{2} \simeq 7.59 \,\times 10^{-4}
        \,\sqrt{\overline{y^2}} \,\Big(\frac{\mathrm{keV}}{m} \Big)\,\Big(\frac{2}{g_d} \Big)^\frac{1}{3}
         \,. \label{Veq}\ee In this expression  $g_d$ is the number of relativistic degrees of freedom at decoupling and
          we introduced the moments\be \overline{y^n} = \frac{\int_0^\infty y^{2+n}\, f_0(y) \, dy}{\int_0^\infty y^2 \,f_0(y)\, dy} \,.\label{avers} \ee

\begin{figure}
\begin{center}
\includegraphics[height=6cm,width=6cm,keepaspectratio=true]{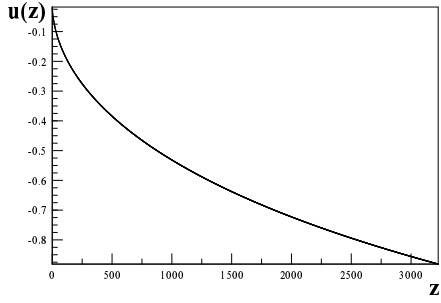}
\caption{$u[z]$ for $z \leq z_{eq}$.} \label{fig:uofz}
\end{center}
\end{figure}

The function $u[z]$ as function of redshift is displayed in fig.
(\ref{fig:uofz}) and \be \ta(u) = \frac{1}{\sinh^2[u]} \,.\label{taofu}\ee

The solution of the Boltzmann equation during stage (III) is given in
ref.(\cite{smallscale}) \bea \tFp(\vk,\vp;s) & = &    -\phi(\vk,s)
\, \Big( p\,\frac{d\tfu(p)}{dp} \Big) + i\,m\,\int_{s_{NR}}^s ds'\,
a^2(s')\, \phi(\vk,s') \Big(\vk\cdot \vec{\nabla}_p \tfu(p) \Big) \,
e^{-i \kpm (s-s')} \nonumber \\  & +  & e^{-i \kpm
(s-s_{NR})}\,\Big[
\tFp(\vk,\vp;\eta_{NR})+\phi(\vk,\eta_{NR})\,\Big(
p\,\frac{d\tfu}{dp} \Big)\Big]\,.  \label{BENR}\eea

The   term $\tFp(\vk,\vp;\eta_{NR})$ in the bracket in eqn. (\ref{BENR}) is the solution of the Boltzmann equation at the beginning of stage (III) (end of
stage (II))  its form is given in detail in ref.\cite{smallscale} but is not necessary in the discussion that follows.

  After radiation-matter equality when the WDM particle is non-relativistic and (DM) perturbations dominate the gravitational potential    and for $k \gg k_{eq}$,  the gravitational potential $\phi$ is determined by Poisson's equation\cite{dodelson} \be \phi (k,\eta) =
-\frac{3}{4}\frac{k^2_{eq}}{k^2\,\ta}\,\td(\vk,s) \label{poissonmat}
\ee  For $s > s_{eq}$, the integral in $s'$ in (\ref{BENR}) is split
from $s_{NR}$ up to $s_{eq}$ and from $s_{eq} $ up to $s$. In the
first integral the gravitational potential is determined by
perturbations in the radiation fluid and in the second integral the
gravitational potential is replaced by Poisson's equation
(\ref{poissonmat}), leading to the result (valid for $s>s_{eq}$)\cite{smallscale}
\bea \tFp(\vk,\vp;s) & = &
\frac{3}{4}\frac{k^2_{eq}}{k^2\,\ta}\,\td(\vk,s) \, \Big(
p\,\frac{d\tfu(p)}{dp} \Big) - i \,\frac{3m k^2_{eq}a^2_{eq}}{4k}
\,\int_{s_{eq}}^s ds'\,      \ta(s')\, \td(k,s') \, \mu\,
\Big(\frac{d \tfu(p)}{dp} \Big) \,  e^{-i \mu Q} \nonumber \\  & +
& \mathcal{F}[\vk,\vp;s]  \,.  \label{delfs}\eea where \be
Q=\frac{kp}{m}(s-s')~;~\mu =
\widehat{\mathbf{k}}\cdot\widehat{\mathbf{p}}\label{Qmudefs}\,,\ee
and  $\mathcal{F}[\vk,\vp;s]$ is given by the second line in (\ref{BENR})
plus the contribution from the integral between $s_{NR} $ and
$s_{eq}$ (for details see ref.\cite{smallscale}).


We are interested in the corrections to the power spectra in the regime of redshift $1 < z \ll z_{eq}$ corresponding to $\ta \gg 1$.

In the asymptotic limit $\ta \gg 1$ when density perturbations
grow as in an Einstein-DeSitter cosmology, $\td \propto \ta$,
therefore in this limit and for  $k \gg k_{eq}$ we can neglect the
first term in (\ref{delfs}). Since in
this limit $\td(k,s') \propto \ta(s') \propto (1/s')^2$ the integral in eqn. (\ref{delfs}) is $\propto 1/s^3$ and dominates all other terms in eqn. (\ref{delfs}) since the last term remains finite in the limit $s \rightarrow 0$\cite{smallscale}.


 Therefore in
the asymptotic limit $\ta \gg 1$ and for small scales $k \gg k_{eq}$
the leading contribution to the perturbation in the distribution
function is given by \be \tFp(\vk,\vp;s) \simeq - i \,\frac{3m
k^2_{eq}a^2_{eq}}{4k} \,\int_{s_{eq}}^s ds'\,      \ta(s')\,
\td(\vk,s') \, \mu\, \Big(\frac{d \tfu(p)}{dp} \Big) \,  e^{-i \mu Q}
\label{fasi} \ee With this form for the distribution function we can
obtain any expectation value once $\td(k,s)$ is determined from the solution of the Boltzmann equation.

         In ref.\cite{smallscale} it is shown that in terms of the variable $u$ defined by eqns. (\ref{sdefd},\ref{udef})  $\delta$ obeys the  fluid-like integro-differential equation
         \bea  && \frac{d^2}{du^2}\td(k,u)-6\,\ta(u)\td(k,u)+ \kappa^2 \td(k,u) -  6\,\alpha  \int_{u_{NR}}^u
  \ta(u')\,\widetilde{\Pi}\big[\alpha(u-u')\big]
\td(k,u')   \,du' \nonumber \\ &  = &   J[k,u]   \label{intdifeq2}
\eea  where the inhomogeneity $J[k,u]$ is given explicitly in
ref.\cite{smallscale},  and  \be
\widetilde{\Pi}\big[\alpha(u-u')\big] = \frac{1}{N}\int_0^\infty y
f_0(y)(\overline{y^2}-y^2) \sin\big[y\,\alpha \,(u-u') \big]
dy~~;~~N = \int_0^\infty y^2 f_0(y)\,dy\,. \label{tilPI}\ee  In the above expressions we introduced
\be \alpha = 2 \sqrt{2}~\frac{k\,T_{0,d}}{m\,k_{eq}a_{eq}} \simeq
2.15 \times 10^{-3}\,
\Big(\frac{k}{k_{eq}}\Big)\,\Big(\frac{2}{g_d}\Big)^{\frac{1}{3}}\,\Big(
\frac{\mathrm{keV}}{m} \Big)\simeq 0.22 \,k
\,\Big(\frac{2}{g_d}\Big)^{\frac{1}{3}}\,\Big(
\frac{\mathrm{keV}}{m} \Big)\times
\big(\mathrm{Mpc}\big)\,,\label{alfa}\ee and \be  \kappa \equiv
\sqrt{\overline{y^2}} ~ \alpha     \,.\label{kapadef} \ee
 In terms of the free streaming wavevector\cite{smallscale}\be k_{fs}=  \frac{\sqrt{3}}{2}\, \frac{k_{eq} }{ \langle \vec{V}^2(t_{eq}) \rangle^\frac{1}{2} } =  \frac{11.17}{\sqrt{\overline{y^2}}} \, \Big( \frac{m}{\mathrm{keV}}\Big)\,
 \Big( \frac{g_d}{\mathrm{2}}\Big)^\frac{1}{3}\,(\mathrm{Mpc})^{-1}\, , \label{keqkfs}\ee
  it follows that \be \kappa = \frac{\sqrt{6}k}{k_{fs}} = \frac{\sqrt{6}\,\lambda_{fs}}{\lambda} \label{kapkfs}\ee where $\lambda_{fs}$ is the free streaming length and $\lambda$ the wavelength of the perturbation.   The (CDM) limit corresponds to $\lambda_{fs} \rightarrow 0$, namely $\kappa \rightarrow 0$, therefore all the WDM corrections are in terms of $\kappa$. 
  
  In the (CDM) limit 
 eqn.(\ref{intdifeq2}) reduces to
Meszaro's equation\cite{mesaros,groth,peeblesbook} for (CDM)
perturbations in a radiation and matter dominated
cosmology\cite{smallscale}. 


   The power spectrum of density perturbations is given by
  \be   P_{\delta}(k) = A k^{n_s}T^2(k) \label{power} \ee where
where $n_s=0.963$\cite{WMAP7} is the index of primordial scalar
       perturbations, $A$ is the amplitude and $T(k)$ is the transfer function. It is convenient to normalize the WDM power spectrum and transfer function to CDM, namely
       \be P_{wdm}(k)= P_{cdm}(k)\, \overline{T}^{\,2}(k)~~;~~\overline{T}(k) = \frac{T_{wdm}(k;\kappa)}{T_{cdm}(k)}
      \label{Tbar}\ee  where
      \be P_{cdm}(k) = A k^{n_s}T^2_{cdm}(k)\label{Pcdm}\ee  is the (CDM) power spectrum and the dependence on WDM is encoded in the $\kappa$ dependence of $T_{wdm}(k;\kappa )$ so that $T_{wdm}(k;\kappa=0)=T_{cdm}(k)$. The dependence on $\kappa$ describes the velocity dispersion and non-vanishing free streaming length of the WDM particle.

    In ref.\cite{smallscale} it is shown that eqn. (\ref{intdifeq2}) can be solved in a systematic Fredholm expansion, from which the transfer function of density perturbations at $z=0$ is extracted. The leading order term is a Born-type approximation which provides a remarkably accurate approximation to the transfer function and reproduces   numerical results available in the literature in several cases (for discussion and comparison see\cite{smallscale}). The definition of the power spectrum and transfer function above are   at $z=0$. We seek to study the redshift dependence for $z \lesssim 30-40$ at which N-body simulations set up initial conditions.


Asymptotically during the matter dominated era as $\ta \rightarrow
\infty$ ($u \rightarrow 0$) it is found\cite{smallscale} that
$\delta(k, u) \rightarrow \ta(u)\, \delta(k,0) +\cdots$ where the
dots stand for subleading terms. The leading and subleading
asymptotic behavior in the $u\rightarrow 0$ ($\ta \rightarrow
\infty$) limit can be obtained from eqn. (\ref{intdifeq2}). In this
limit the inhomogeneity $J[k,0]$ is a finite constant (see
expressions in ref.\cite{smallscale}), the integral term receives
the largest contribution for $u' \sim u \sim 0$ and in this region
we find \be -6\alpha \widetilde{\Pi}\big[\alpha(u-u')\big] \simeq
\kappa^4 (u-u')^3 \Big(1-\beta_2\Big)+\cdots\,. \label{Piappx}\ee
where \be \beta_2 = \frac{\overline{y^4}}{\big(\overline{y^2}
\big)^2} \label{kurtosis}\ee is the \emph{kurtosis} of the
distribution function of the decoupled particle,  with the moments
  defined by eqn. (\ref{avers}), and the dots stand for terms that yield subleading corrections (see below).

 Since
 \be \ta(u) = \frac{1}{\sinh^2[u]} \sim \frac{1}{u^2} - \frac{1}{3} +  \mathcal{O}(u^2)
 \,,\label{tauasi}\ee  we propose the asymptotic expansion
 \be \delta(k,u) = \frac{\delta(k,0)}{u^2} + \delta_1(k) + \delta_2(k)~u^2 \,\ln[-u]
 +\cdots \, \label{delexp}\ee Introducing this expansion in eqn. (\ref{intdifeq2})
 we find \be \delta_1(k) = \frac{\delta(k,0)}{6}\big[\kappa^2+2\big] ~~;~~ \delta_2(k) = -\delta(k,0)\,
 \frac{ \kappa^4}{4}\Big(1 - \beta_2\Big)\,,\label{coeffs}\ee
  where $\delta(\vk,0)$ is obtained from the asymptotic solution of the full eqn. (\ref{intdifeq2}). Therefore
  \be \delta(\vk,u)= \delta(\vk,0)\,D[k,u] \label{capD}\ee where    the \emph{wavevector dependent} growth
  factor is found to be \be D[k,u] = D_{cdm}[u]~\overline{D}[k,u] \label{growthfacdef}\ee with $D_{cdm}[u]$ being the CDM
  growth factor (for $\kappa =0$) \be D_{cdm} =  \frac{1}{u^2}\,\Bigg[1+\frac{u^2}{3}+\cdots\Bigg]  \label{Dcdm}\ee and \be \overline{D}[k,u] =  \Bigg[1+ \frac{(\kappa\,u)^2}{6}+\frac{(\kappa\,u)^4}{4}(-\ln[-u])\,\Big(1 - \beta_2\Big)+\cdots\Bigg]\,. \label{delofkuwdm}\ee  contains the WDM corrections as is manifest in the $\kappa$ dependence.

  For $u \rightarrow 0$ we find
  $$D_{cdm}[u] = \frac{1}{u^2} +\frac{1}{3} \simeq \ta +\frac{2}{3}$$
  which is recognized as the growing solution of Meszaros equation for CDM\cite{mesaros,groth,peeblesbook}. Furthermore from (\ref{sdefd})  we recognize that \be -\kappa\,u = k \,l_{fs}\big[\frac{\sqrt{\langle p^2 \rangle}}{m},\eta_0,\eta\big] \label{prodkapau}\ee where $l_{fs}\big[\frac{\sqrt{\langle p^2\rangle}}{m},\eta_0,\eta\big]$ is the comoving free streaming distance that a particle with (comoving) velocity $\sqrt{\langle p^2 \rangle}/m$ travels between conformal time $\eta$ and today $\eta_0 \gg 1$. We see that up to logarithms, the expansion in powers of $\kappa\,u$ is valid at late times for wavelengths much larger than the free streaming distance that the particle would travel between that time and today.

 The identification (\ref{prodkapau})  leads to a simple physical interpretation of  the first term in  $\overline{D}[k,u]$: free streaming of collisionless particles suppresses the gravitational collapse of density perturbations, the longer
  the time scale, the farther   the free streaming particles can travel away from the collapsing
  region erasing the perturbations. Therefore the first term reflects that at earlier times (larger values of $u$) density perturbations are \textit{larger}. The second term, however, has a more interesting interpretation. As it will be discussed below, it represents the peculiar velocity contribution to free streaming induced by
  gravitational self-interaction (see discussion on peculiar velocity below). When  $\beta_2 > 1$ the peculiar velocity contribution \emph{increases} the free streaming velocity leading to
  a \textit{suppression} of power, which counterbalances the enhancement by the first term. Which term dominates depends
    on the scale $k$, the free streaming wavevector, a characteristic of the WDM particle,  and the
    redshift. This will be analyzed in two specific models below.

      We emphasize that the expansion in (\ref{delexp}) is
  valid at long time, in particular for $\kappa\,u < 1$. At higher orders in the expansion, the terms that
  feature the $\ln(-u)$ only appear \emph{linearly } in the logarithm but multiplied by higher powers of
  $\kappa \,u$, therefore for $|\kappa\,u| < 1$ the third term in $\overline{D}$ is  the
  \emph{leading logarithmic} contribution, with higher  contributions being of the form
  $(\kappa\,u)^n\,\ln(-u)~;~n = 6, 8\cdots$. This is an important observation:
  in particular within the regime of validity of the perturbative
  expansion $|\kappa u| < 1$ it is still possible that $|\kappa u \,\ln(-u)| \sim 1$
   and the second term in (\ref{delofkuwdm}) can balance
  the first term within the region of validity of the approximation.

   An estimate of the range of validity is obtained from  \be 0.098 \leq -u[z] \leq 0.125 ~~for~~ 30 \leq z \leq 50 \,, \label{urange}\ee for example in the region of redshifts where initial conditions for N-body simulations are set up, the WDM corrections to the growth factor are of $\mathcal{O}(10-15\%)$ for $k \gtrsim (1- 2)\, k_{fs}$ which for a species with $m \sim \mathrm{keV}$  decoupled with $g_d \sim 30-100$ with   $\overline{y^2} \sim 10$ corresponds to $k  \gtrsim 10-30 \,(\mathrm{Mpc})^{-1}$.


   There is a caveat in this  analysis of the reliability of the expansion, since it applies \emph{only in the linear regime} where the linearized Boltzmann equation describes the transfer function. It is conceivable that non-linear effects restrict further the regime of validity, but of course this cannot be assessed in the linear theory which is the focus of this discussion.


   Using Poisson's equation (\ref{poissonmat}),  the asymptotic behavior
   $\td(\vk,u) \rightarrow \td(\vk,0)\,\ta(u)$ and the definition of the transfer function\cite{dodelson}
   $T(k)$ \be \phi(k,\ta\gg 1) = \frac{9}{10}\phi_i(k) T(k) \label{Tofkdef}\ee where
   $\phi_i(k)$ is the primordial value of gravitational perturbations seeded by inflation.
   It then follows that \be \td(\vk,0) = -  \phi_i(\vk)\, \frac{6\, k^2}{5  \,k^2_{eq}} T(k)\,. \label{delexact}\ee

We emphasize that there are \emph{two} different averages: i) the \emph{statistical} average of a quantity $\mathcal{O}$ with the
   perturbed distribution function $f_0 + F_1$ to which we  refer as $\langle \mathcal{O} \rangle$, ii) average over  the initial gravitational potential $\phi_i$ which is a stochastic Gaussian field (we neglect
    possible non-Gaussianity) whose power spectrum is determined during the inflationary era
    \be  \overline{\phi_i(\vk)\phi_i(-\vk')}= (2\pi)^3\,\delta^{(3)}(\vk-\vk')\,P_\phi(k)
     \label{powerspecphi}\ee
   where the $\overline{A\,B}$ refers to averages with the primordial Gaussian distribution
   function for the gravitational potential\footnote{This definition should \emph{not} be confused with that
   of the moments in eqn. (\ref{avers}) which refer to averages with the unperturbed distribution function.
    The meaning
   of averages is unambiguously inferred from the context.}. Therefore \emph{full expectation values}
    correspond to averages both with the perturbed distribution function and the Gaussian
     distribution function for the primordial gravitational potential, these are given by
      $\overline{\langle \mathcal{O} \rangle}$, with the power spectrum of
      \emph{matter density fluctuations} \be\overline{\delta(\vk,0)\delta(-\vk',0)}=
      (2\pi)^3\,\delta^{(3)}(\vk-\vk')\,P_\delta(k) \,,
      \label{powerdelta}\ee  where $P_\delta(k)$ is given by eqn. (\ref{power}).

   Including the wavevector dependent growth factor $\overline{D}[k,u]$ (\ref{delofkuwdm}) but keeping only the (WDM) ($\kappa \neq 0$) corrections with redshift,    the \emph{effective}  (WDM)  power spectrum at
     $z\ll z_{eq}$ is given by  \be P_\delta[k,z]  = P_{wdm}(k)\,\overline{D}^{\,2}[k,z]
      \label{powofz}\ee where   the scale and redshift dependent correction is given by (see eqn. (\ref{delofkuwdm}))

      \be \overline{D}[k,z] = 1+
      \frac{\kappa^2}{6}\,\Bigg[\frac{1+z}{1+z_{eq}}\Bigg]-
       \frac{\kappa^4}{8}\,\Bigg[\frac{1+z}{1+z_{eq}}\Bigg]^2 \,\ln\Bigg[\frac{1+z_{eq}}{1+z}\Bigg]\,
       \Big(  \beta_2-1\Big)+\cdots \label{overDofz}  \ee

      For $\beta_2 > 1$ the third term is negative and competes with the second term, dominating the
      corrections for scales \be k > k_{fs} \,\Bigg[ \frac{2\,(1+z_{eq})}{9\,(1+z )(\beta_2-1) \ln\Big[\frac{(1+z_{eq})}{(1+z)} \Big]} \Bigg]^\frac{1}{2} \ee for $\beta_2 -1 \sim \mathcal{O}(1)$ and $z \simeq 30-50$ one finds that the third term dominates over the second for $k \sim (1-2) \,k_{fs}$. These are the scales beyond which   the contribution from the peculiar velocities to free streaming leads to       a \emph{suppression} of the power spectrum. Coincidentally this is the scale at which the power spectrum displays (WDM) acoustic oscillations which arise from the competition between free streaming and gravitational collapse in the \emph{collisionless} regime as described in ref.\cite{smallscale}.

\section{Peculiar velocity and phase space density:}\label{sec:velocity}

         \emph{Statistical} averages of observables with the perturbed distribution function (\ref{f}) in the linearized theory (in terms of their spatial Fourier transform) are given by \be   \widetilde{ {\mathcal{O}}} (\vec{k};\eta)     \equiv  \langle \mathcal{O}(\vp,\vk,\eta) \rangle =  \frac{\int \frac{d^3p}{(2\pi)^3} \Big[f_0(p)+F_1(\vp,\vk;\eta)\Big]\,\mathcal{O}(\vp,\vk;\eta) }{\int\frac{d^3p}{(2\pi)^3} \Big[f_0(p)+F_1(\vp,\vk;\eta)\Big]   } = \frac{  \widetilde{\mathcal{O}}_0( \vk;\eta) +   \Delta\widetilde{ \mathcal{O}} ( \vk;\eta) }{\Big[1+\delta(\vk,\eta)\Big]}   \label{expvals}\ee  where
\bea  \widetilde{\mathcal{O}}_0(\vk;\eta) & = &
 {\int\frac{d^3p}{(2\pi)^3}\tfu(p) \,\mathcal{O}(\vp,\vk;\eta)} \,, \label{expval0}\\
\Delta\widetilde{\mathcal{O}} (\vk;\eta) & = & \int
\frac{d^3p}{(2\pi)^3} \, \tFp(\vp,\vk;\eta)\,
\mathcal{O}(\vp,\vk;\eta)\,. \label{expval1stord}\eea where
$\tFp,\tfu$ are defined in eqn. (\ref{tildes}). In the linearized
approximation
 \be    \widetilde{\mathcal{O}}(\vec{k};\eta)  \simeq
  \widetilde{\mathcal{O}}_0(\vk;\eta)  + \Big(
   \Delta\widetilde{\mathcal{O}} ( \vk;\eta) -\widetilde{\mathcal{O}}_0(\vk;\eta)\,  \delta(\vk,\eta) \Big)  \,. \label{linearexpval}\ee
    With $\tFp(\vk,\vp;s)$ given by (\ref{fasi}) and $\delta(k,\eta)$ given by (\ref{delexp},\ref{coeffs}) we can now obtain
    any statistical average by expanding $\mathcal{O}(\vp,\vk;\eta) \equiv
     \mathcal{O}(p,k,\mu;\eta)$ in Legendre polynomials in $\mu$ and carrying out the integrals
     in $p,\mu$ leading to an expansion in spherical Bessel functions. However,  here we focus on obtaining the leading asymptotic expansion of these averages for $z \ll z_{eq}$, namely for $u \ll 1$. This is readily achieved by using the asymptotic expansion (\ref{delexp})  with the coefficients given by (\ref{coeffs}),
expanding

$$\exp[-i\mu Q]\simeq 1- i\mu Q -\frac{1}{2}\mu^2 Q^2 + \frac{i}{6}\mu^3 Q^3 +\cdots $$

\noindent and integrating over $\mu$ and $p$ term by term in the expansion.

\subsection{Peculiar velocity}
Writing the comoving peculiar velocity in terms of the longitudinal
and transverse components
 \be \vec{v}(\vk,\eta) =  \langle \frac{\vp}{m} \rangle  \equiv \vec{v}_T +
 \widehat{\mathbf{k}}\,v_L ~~,~~ \vec{k}\cdot\vec{v}_T =0 \label{velcom}  \ee
 where \be  {v}_L =   \frac{p}{m}\,\mu ~~;~~ \mu = \widehat{\mathbf{k}}\cdot \widehat{\mathbf{p}}
   \label{vpara}\ee and $p$ is   the comoving momentum. In the linearized approximation,
    the expectation value of $ k \,v_L$  is given by \be k\, v_L(\vk,\eta)    =
     \int \frac{d^3p}{(2\pi)^3} \tFp(\vp,\vk;\eta)\,\frac{\vk\cdot \vp}{m} \,.\label{expvL}\ee
     Furthermore,  $\tFp(\vp,\vk;\eta)$ is a function of $k$ and $\vk\cdot\vp$ leading to $\vec{v}_T =0$ in the linearized approximation.  Since the gravitational potential is only a function of $k$, the first term on the right hand side of  (\ref{BENR}) does not contribute and we find  \be
k \,  v_L(\vk,\eta)   = i\,\frac{d}{ds} \int
\frac{d^3p}{(2\pi)^3}\Bigg[ \tFp(\vp,\vk;s)+\phi(k,s) \Big(p
\frac{d\tfu(p)}{dp} \Big) \Bigg] = i\frac{d}{ds}
\Bigg[\delta(\vk,s)-3 \phi(k,s)\Bigg] \label{expvL2}\ee Using $d/ds
= a d/d\eta$ equation (\ref{expvL2}) becomes  \be
\frac{d\delta}{d\eta} - 3 \frac{d\phi}{d\eta} + i\frac{k}{a} \,
v_L   =0 \label{conteqn}\ee which is recognized as the
continuity equation in presence of the gravitational
potential\cite{dodelson}\footnote{Note that the Newtonian potential
in eqn. (\ref{gij}) features a minus sign with respect to the
definition in \cite{dodelson}.} for the \emph{comoving} longitudinal
velocity. For $\ta \gg 1$ and $k\gg k_{eq}$ the second term in the
continuity eqn. (\ref{conteqn}) can be safely neglected, leading to
\be   v_L(\vk,u)   = i\,\frac{k_{eq}a_{eq}}{2\sqrt{2}\,k}
\frac{d\delta(\vk,u)}{du}\,.
\label{consolvL}\ee  As a function of redshift we find
\be v_L(\vk,z) =   i\,\frac{k_{eq}a_{eq}}{ \sqrt{2}\,k}
\,\frac{\delta(k,0)}{(-u[z])^3}\,\mathcal{V}_{wdm}[k,z]\label{vLofkz}\ee
where we used  the asymptotic expansion
(\ref{delexp},\ref{coeffs}) and introduced \be
\mathcal{V}_{wdm}[k,z] = \Bigg[1+
\frac{ \kappa^4}{8}\Big[\frac{1+z}{1+z_{eq}} \Big]^2\, \ln\Big[\frac{1+z_{eq}}{1+z}\Big] \,\Big (
\beta_2-1 \Big)  +\cdots
\Bigg]\,.  \label{bigTau}\ee

 In the $CDM$ limit $\kappa \rightarrow 0 $ the growth factor $u^{-3} \simeq \ta^{\frac{3}{2}}$ which is recognized as the growth of \emph{comoving} peculiar velocity in a matter dominated cosmology, in this limit $\overline{T}(k)\rightarrow 1$\cite{smallscale} and
  $\mathcal{V}_{wdm}[k,u]\rightarrow 1$.  The
 function $\mathcal{V}_{wdm}[k,u]$   encodes  the corrections to the peculiar velocity at small scales.
  It is clear that as compared to the CDM case, when the kurtosis $\beta_2 > 1$ the
  peculiar velocity  at small scales $\kappa  \gtrsim 1 $ is \emph{larger} at higher
  redshift. Comparing eqn. (\ref{bigTau}) with the third term in eqn.
  (\ref{overDofz})     confirms the interpretation of the suppression of the power spectrum
  at small scales and high redshift as a consequence of the  peculiar velocity contribution   to free streaming.

\subsection{Statistical fluctuations and correlation functions:}

In the linearized approximation (and with   adiabatic
perturbations only), the perturbation in the distribution function
$\tFp(\vk,\vp;u)$ is linear in the primordial gravitational
potential $\phi_i(k)$ which is a Gaussian variable determined by the
power spectrum of perturbations during the inflationary stage (here
we neglect possible non-gaussianities).
 Therefore as discussed in the previous section there are two different averages, i)
  a \emph{statistical} average with the perturbed
   distribution function $f_0+F_1$,  ii) with the initial
  Gaussian probability distribution of $P_\phi(k)$ in eqn. (\ref{powerspecphi}).

\emph{Statistical} fluctuations are contained in the variance of the
various quantities calculated with the perturbed distribution
$\tFp$. These are   linear in $\delta(k,0)$, namely linear in
$\phi_i$, therefore they feature Gaussian fluctuations with the probability
distribution function $P_{\phi}(k)$, but with non-gaussian
\emph{statistical} variances.

 As an  example of a statistical
fluctuation consider

  \be \Delta   v^2_L   = {6}  \langle
 \frac{ p^2}{m^2}\rangle_0 \, \int^u du' \ta(u')\td(k,u') \Bigg[ (u-u') -
  \frac{\kappa^2}{6}\,\beta_2 \, (u-u')^3 \cdots \Bigg]\,du' \label{vL2}\ee Using the
  asymptotic expansions (\ref{tauasi},\ref{delexp}),\ref{coeffs}) we
  find up to leading logarithmic order
  \bea \int^u \, \ta(u')\,\td(k,u')   (u-u')\,du' & = &
  \frac{\delta(k,0)}{6\,u^2} \Bigg[1 - \kappa^2\,u^2\,\ln[-u]\Bigg]
  +\cdots \label{linearterm} \\
  \int^u \, \ta(u')\,\td(k,u')   (u-u')^3 \,du' & = &
  -\delta(k,0)  \,\ln[-u]
  +\cdots \label{cubicterm}\eea leading to similar statistical fluctuations for the total velocity
  dispersion $\langle p^2/m^2 \rangle$  and the transverse component $\vec{v}_T$, namely (all quantities are comoving)
  \be \Delta \langle v^2_L \rangle = \frac{\langle
  p^2\rangle_0}{m^2}   \, \frac{\delta(k,0)}{u^2} \Bigg[1 -\kappa^2\,u^2
  \,\ln[-u]\Big(1- \beta_2\Big) \Bigg] +\cdots \label{delvL2fin}\ee

   \be \Delta \langle \frac{p^2}{m^2} \rangle = \frac{5}{3}\,\langle\frac{ p^2}{m^2} \rangle_0 \,
   \frac{\delta(k,0)}{u^2} \Bigg[1  -\kappa^2\,u^2
  \ln[-u]\Big( 1- \frac{21}{25}\,\beta_2\Big)\Bigg]+\cdots \label{deltap2}\ee

  \be \Delta \langle v^2_T \rangle = \frac{2}{3} \,\frac{\langle
  p^2\rangle_0}{m^2}   \,  \frac{\delta(k,0)}{u^2}  \Bigg[1 -\kappa^2\,u^2
  \,\ln[-u]\Big(1- \frac{3}{5}\beta_2\Big) \Bigg] +\cdots \label{delvT2fin}\ee Restoring
  units, writing   $\langle p^2 \rangle_0 = \overline{y^2}\,T^2_{0,d}$ and expressing these expressions in terms of redshift, we find the following \emph{statistical fluctuations}
  \be \Delta \langle v^2_L \rangle \simeq  16.33\, \Big(\frac{\mathrm{km}}{\mathrm{sec}}\Big)^2 \, \Big(\frac{\mathrm{keV}}{m} \Big)^2 \, \overline{y^2}\, {\delta(k,0)} \, \Big[\frac{1+z_{eq}}{1+z}\Big]~\Bigg[1 -\kappa^2\,\Big[\frac{1+z}{1+z_{eq}}\Big]
  \,\ln\Big[\frac{1+z_{eq}}{1+z}\Big]\Big(\beta_2-1\Big) \Bigg] +\cdots \label{delvL2finofz}\ee

   \be \Delta \langle \frac{p^2}{m^2} \rangle \simeq 27.22\,\Big(\frac{\mathrm{km}}{\mathrm{sec}}\Big)^2 \, \Big(\frac{\mathrm{keV}}{m} \Big)^2 \,  \overline{y^2}\,{\delta(k,0)} \, \Big[\frac{1+z_{eq}}{1+z}\Big]~\Bigg[1 -\frac{21}{25}\,\kappa^2\,\Big[\frac{1+z}{1+z_{eq}}\Big]
  \,\ln\Big[\frac{1+z_{eq}}{1+z}\Big]\Big( \beta_2-\frac{25}{21}\Big) \Bigg] +\cdots  \label{deltap2ofz}\ee

  \be \Delta \langle v^2_T \rangle \simeq 10.89 \,\Big(\frac{\mathrm{km}}{\mathrm{sec}}\Big)^2 \,\Big(\frac{\mathrm{keV}}{m} \Big)^2 \, \overline{y^2}\, {\delta(k,0)} \, \Big[\frac{1+z_{eq}}{1+z}\Big]~\Bigg[1 -\frac{3}{5}\kappa^2\,\Big[\frac{1+z}{1+z_{eq}}\Big]
  \,\ln\Big[\frac{1+z_{eq}}{1+z}\Big]\Big(\beta_2-\frac{5}{3}\Big) \Bigg] +\cdots \label{delvT2finofz}\ee

  These expressions also show that the WDM corrections (proportional to $\kappa^2$) \emph{suppress}
   the statistical fluctuations at   small scales $k  \gtrsim k_{fs}$ where the peculiar velocity contribution to  free streaming becomes important (we will see below that at least for the  (WDM) candidates considered here $\beta_2 > 2$).

  The peculiar velocity autocorrelation function is given by \be \xi_{ij}(\vec{x},\vec{x'};u) = \int \frac{d^3k}{(2\pi)^3} e^{i\vk\cdot\vec{x}} \int \frac{d^3k'}{(2\pi)^3} e^{-i\vk'\cdot\vec{x}'}~ \overline{v_{i }(\vk,u)\,v^*_{j }(-\vk',u)} \label{correvel}\ee using (\ref{powerdelta}) and (\ref{consolvL}) we find
  \be \xi_{ij}(\vec{r};z) = \frac{k^2_{eq}a^2_{eq}}{2\,u^6}\,\int \frac{d^3k}{(2\pi)^3} e^{i\vk\cdot\vec{r}} \, \widehat{k}_i \widehat{k}_j \,\frac{ {P}_{\delta}(k)}{k^2}\,\mathcal{V}_{wdm}[k,z] ~~;~~ \vec{r}=\vec{x}-\vec{x'}\,.\label{velcorrfin} \ee Since there is only one vector $\vec{r}$ we write \be \xi_{ij}(\vec{r},z) = \mathcal{P}^{\perp}_{ij}(\widehat{r})\,\xi^{\perp}(r;z)+ \mathcal{P}^{\parallel}_{ij}(\widehat{r})\,\xi^{\parallel}(r;z) \label{corrs}\ee where \be  \mathcal{P}^{\perp}_{ij}(\widehat{r})  = \delta_{ij}-\widehat{r}_i\,\widehat{r}_j~~;~~ \mathcal{P}^{\parallel}_{ij}(\widehat{r})   =   \widehat{r}_i\,\widehat{r}_j \,.\label{projectors}\ee are the projectors on directions parallel and perpendicular to $ {\bf{r}}$.

  We find
  \bea \xi^{\parallel}(r;z) & = & \frac{k^2_{eq}a^2_{eq}}{12\,\pi^2\,\,(u[z])^6}\,\int  {dk}  \, {P_{\delta}(k)} \,\mathcal{V}_{wdm}[k,z]\,\Big[j_0(kr)-2\,j_2(kr)\Big] \label{xipara}\\
  \xi^{\perp}(r;z) & = & \frac{k^2_{eq}a^2_{eq}}{6\,\pi^2\,(u[z])^6}\,\int  {dk}  \, {P_{\delta}(k)} \,\mathcal{V}_{wdm}[k,z]\,\Big[j_0(kr)+\,j_2(kr)\Big]
  \eea where $j_{0,2}$ are spherical Bessel functions.

  Thus we see that the \emph{effective} (WDM) power spectrum for peculiar velocities is ${P_{\delta}(k)} \,\mathcal{V}_{wdm}[k,z]$.

  From expression (\ref{bigTau}) it is clear that for $\beta_2 >1$ (WDM) perturbations \emph{enhance} the peculiar velocity autocorrelation function for $z \simeq 30-50$. This enhancement of the velocity correlation function is in concordance with the suppression of the power spectrum, since the larger velocity dispersion induced by self-gravity leads to a larger free streaming velocity and a further suppression of the power spectrum.

  \subsection{Phase space density:}

  In their seminal article Tremaine and Gunn \cite{TG} argued that the
coarse grained phase space density is always smaller than  or equal
to the maximum of the (fine grained) microscopic phase space
density, namely, the distribution function, allowing to establish bounds on the
mass of the DM particle.

Such argument relies on
a theorem \cite{theo,dehnen} that states that collisionless phase
mixing or violent relaxation by gravitational dynamics (mergers or accretion) can only
diminish the coarse grained phase space density. A similar argument
was presented in refs. \cite{dalcanton1,hogan,coldmatter,darkmatter,hectornorma} where a \emph{proxy} for a coarse-grained phase space density in absence
of gravitational perturbations was introduced.

 However, whereas the distribution function obeys the collisionless
Boltzmann equation, Dehnen\cite{dehnen} clarifies that the
coarse grained phase space density does not necessarily evolve with the collisionless Boltzmann equation, and introduces an \emph{excess mass function} which is argued to always diminish upon gravitational phase space mixing.

 Numerical simulations confirm the evolution of a coarse grained phase space density
towards smaller values during violent relaxation events such as encounters, mergers and accretion of haloes\cite{Qpeirani,Qhoffman}.
In the simulations in ref.\cite{Qpeirani} a phase space density $Q$ is obtained by averaging $\rho,\sigma$ over a determined volume, and its evolution with redshift is
followed from $z=10$ until $z=0$ diminishing by a factor $\simeq 40$ during this interval.

However, to the best of our knowledge, a consistent study of the evolution of the microscopic phase space density
including gravitational effects even in the linearized approximation has not yet been provided.

In linearized theory, the corrections to the distribution function $F_1$, or rather the normalized perturbation $\tFp(\vp,\vk,\eta)$ defined by eqn. (\ref{tildes}) obeys the collisionless Boltzmann
equation (\ref{BE}), whose solution in the regime when the DM is non-relativistic is given by eqn. (\ref{BENR}). Thus the time evolution of the \emph{microscopic} phase space density is completely
determined. Two aspects of this solution invite further scrutiny: i) density perturbations \emph{grow} from
self-gravity effects, ii) peculiar velocities \emph{also grow}, a direct consequence of the continuity
equation (\ref{conteqn}) and explicitly shown by eqn. (\ref{vLofkz}). That both quantities \emph{grow} upon gravitational collapse suggests an examination of the phase space evolution in the linearized regime.

In principle one could perform the Fourier transform back to (comoving) spatial coordinates and
obtain $\tFp(\vp,\vec{r};\eta)$, however, $\delta(\vk,\eta)$ is a \emph{stochastic variable} with a Gaussian probability distribution determined by the power spectrum of the primordial gravitational potential. Therefore, the linear correction to the microscopic phase space density itself becomes a stochastic variable as discussed
above.

 Rather than pursuing the Fourier transform, which in the linearized approximation can be performed at
 any state in the  calculation, we follow refs.\cite{dalcanton1,hogan,coldmatter,darkmatter,hectornorma,boysnudm} and
define the coarse grained (dimensionless) primordial phase space
density \be \mathcal{D}  \equiv \frac{n(t)}{\big\langle \vec{P}^2_f
\big\rangle^\frac32} \; , \label{D} \ee   where $   \vec{P}^2_f   =   \vec{p}  /a(t)  $ is the
\emph{physical momentum}.  In absence of gravitational perturbations,  the (unperturbed) distribution function of the decoupled species is frozen and $n(t)=n_0/a^3(t)$, therefore  it is clear that $ \mathcal{D} $ is a
\emph{constant}, namely a Liouville invariant. In absence of
self-gravity  it is given by   \be \displaystyle \mathcal{D}_0 = \frac{
g}{2\pi^2} \frac{\Bigg[\int_0^\infty y^2 f_0(y) dy
\Bigg]^{\frac{5}{2}}}{\Bigg[\int_0^\infty y^4 f_0(y)dy
\Bigg]^{\frac{3}{2}}}\; , \label{Dex} \ee where $ f_0(y) $ is the
decoupled distribution function, and $g$ the number of internal
degrees of freedom of the WDM particle.

When the particle becomes non-relativistic $ \rho(t) = m  \; n(t) $
and $ \big\langle \vec{V}^2 \big\rangle = \big\langle
\frac{\vec{P}^2_f}{m^2} \big\rangle $, therefore, \be \mathcal{D}  =
\frac{\rho}{m^4\,\big\langle \vec{V}^2 \big\rangle^{\frac{3}{2}} } =
\frac{Q_{DH}}{m^4}\label{DNR} \ee where $ Q_{DH}= {\rho}/\big\langle
\vec{V}^2 \big\rangle^{\frac32}  $ is the phase-space density
introduced in refs. \cite{dalcanton1,hogan}.

In the non-relativistic regime $ \mathcal{D} $ is related to the
coarse grained phase space density $ Q_{TG} $ introduced by Tremaine
and Gunn \cite{TG} \be \label{QTG} Q_{TG}= \frac{\rho}{m^4 \; (2 \,
\pi \; \sigma^2)^\frac32} =  \left( \frac3{2 \, \pi} \right)^\frac32
\; \mathcal{D}  \; . \ee where $\sigma$ is the one-dimensional
velocity dispersion. The observationally accessible quantity is the
phase space density $ \rho/\sigma^3 $, therefore, using  $ \rho = m
   n $  for a decoupled particle that is non-relativistic today and
eq.(\ref{DNR}), we define the primordial phase space
density\footnote{  $ \mathrm{keV}^4 \;
\big(\mathrm{km}/\mathrm{s}\big)^3 = 1.2723 \; 10^8 \;
\frac{M_\odot}{\mathrm{kpc}^3} $.}

\be \frac{\rho_{DM}}{\sigma^3_{DM}} = 3^\frac32 \;  m^4  \;
\mathcal{D} \equiv 6.611 \times 10^8  \;  \mathcal{D} \;
\Big[\frac{m}{\mathrm{keV}}\Big]^4  \;
\frac{M_\odot/\mathrm{kpc}^3}{\big(\mathrm{km}/\mathrm{s}\big)^3}
 \; \,. \label{PSDM}
\ee

In refs.\cite{coldmatter,darkmatter,boysnudm,hectornorma} the phase mixing theorem was invoked to argue that the \emph{observed} phase space density is smaller than the primordial value (\ref{PSDM}) with $\mathcal{D}$ replaced by $\mathcal{D}_0$ given by eqn. (\ref{Dex}), leading to lower bound on the mass of the (WDM) particle. However, as emphasized in ref.\cite{dehnen} the phase mixing theorem\cite{theo} does not directly address the evolution of $\mathcal{D}$, nor has there yet been an analysis of its evolution in the linearized regime.

The results obtained above allow us to directly calculate the corrections to $\mathcal{D}$ from self-gravity in the linearized theory. Using the identity (\ref{expval1stord}) for linearized statistical averages,  it is given by

  \be \mathcal{D} = \frac{n_0\Big[1+\delta(k,u) \Big]^\frac{5}{2}}{\Bigg[\langle p^2 \rangle_0+\Delta\langle p^2
   \rangle\Bigg]^{\frac{3}{2}}} \simeq \mathcal{D}_0\,\Bigg[1  +  \frac{21}{20}
\, \delta(\vk,0) \, {\kappa^2} \,  \ln\Big[ \frac{1+z_{eq}}{1+z}
\Big]\,  \Big(  \beta_2 -\frac{25}{21}\Big)  \Bigg]
\,.\label{psdens}\ee were we have used
eqns.(\ref{delexp},\ref{coeffs}) and (\ref{deltap2}). In the (CDM)
limit $\kappa \rightarrow 0$ this coarse grained phase space density
remains constant at least up to linear order in gravitational
perturbations. However, for (WDM) eqn. (\ref{psdens}) clearly
indicates that in the regions where matter density perturbations are
positive the phase space density \emph{increases} with the logarithm
of the scale factor when $\beta_2 > 1.19$. We will see below that
this the case at least for two examples of (WDM) candidates
supported by particle physics models. The reason for the increase in
the coarse grained phase space density can be tracked to the
suppression of statistical fluctuations: the leading term in
$\delta(\vk,u) = \delta(\vk,0)/u^2 +\cdots$ cancels against the
leading term proportional to $\delta(\vk,0)/u^2$ in the statistical
fluctuation (\ref{deltap2}), these are the only contributions that
remain in the (CDM) limit, however the (WDM) contribution suppresses
the statistical fluctuation of the velocity dispersion leading to an
\emph{increase} of the coarse  grained phase space density as a consequence of the suppression of the statistical fluctuations (statistical variance) of the velocity dispersion in the (WDM) case.

\section{Two specific examples.}

We now focus on two specific examples of WDM candidates: sterile neutrinos produced via the Dodelson Widrow (DW) mechanism\cite{dw}
for which \be f_{dw}(y) = \frac{\xi}{e^y+1}\label{fdw}\ee where the constant $\xi$   is a function of the active-sterile mixing angle\cite{dw}, and sterile neutrinos produced near the electroweak scale via the decay of scalar or vector bosons (BD) for which\cite{boysnudm,jun}
 \be f_{bd}(y) = \frac{\lambda}{\sqrt{y}}\,\sum_{n=1}^\infty \frac{e^{-ny}}{n^\frac{5}{2}} \label{fbd}\ee where $\lambda \sim 10^{-2}$.

  We implement the Born approximation to the matter power spectrum presented in ref.\cite{smallscale} to obtain the corrected power spectrum
 normalized to (CDM) $[\overline{T}(k)\overline{D}[k,z]]^2$. As discussed in ref.\cite{smallscale} the Born approximation yields excellent agreement with the
 power spectrum obtained in ref.\cite{vieldwdm} for (DW) sterile neutrinos.

 \vspace{2mm}

 \textbf{(DW) sterile neutrinos:}

 \vspace{2mm}

 For the  distribution function (\ref{fdw}) we find: \be \overline{y^2} = 12.939 ~~;~~ \beta_2 = 2.367 ~~;~~k_{fs} = 5.44 \, \Big( \frac{m}{\mathrm{keV}}\Big)\,\Big( \frac{g_d}{10.75}\Big)^\frac{1}{3} \,\Big(\mathrm{Mpc} \Big)^{-1} \label{overysdw}\ee

\begin{figure}[ht]
\begin{center}
\includegraphics[height=8cm,width=8cm,keepaspectratio=true]{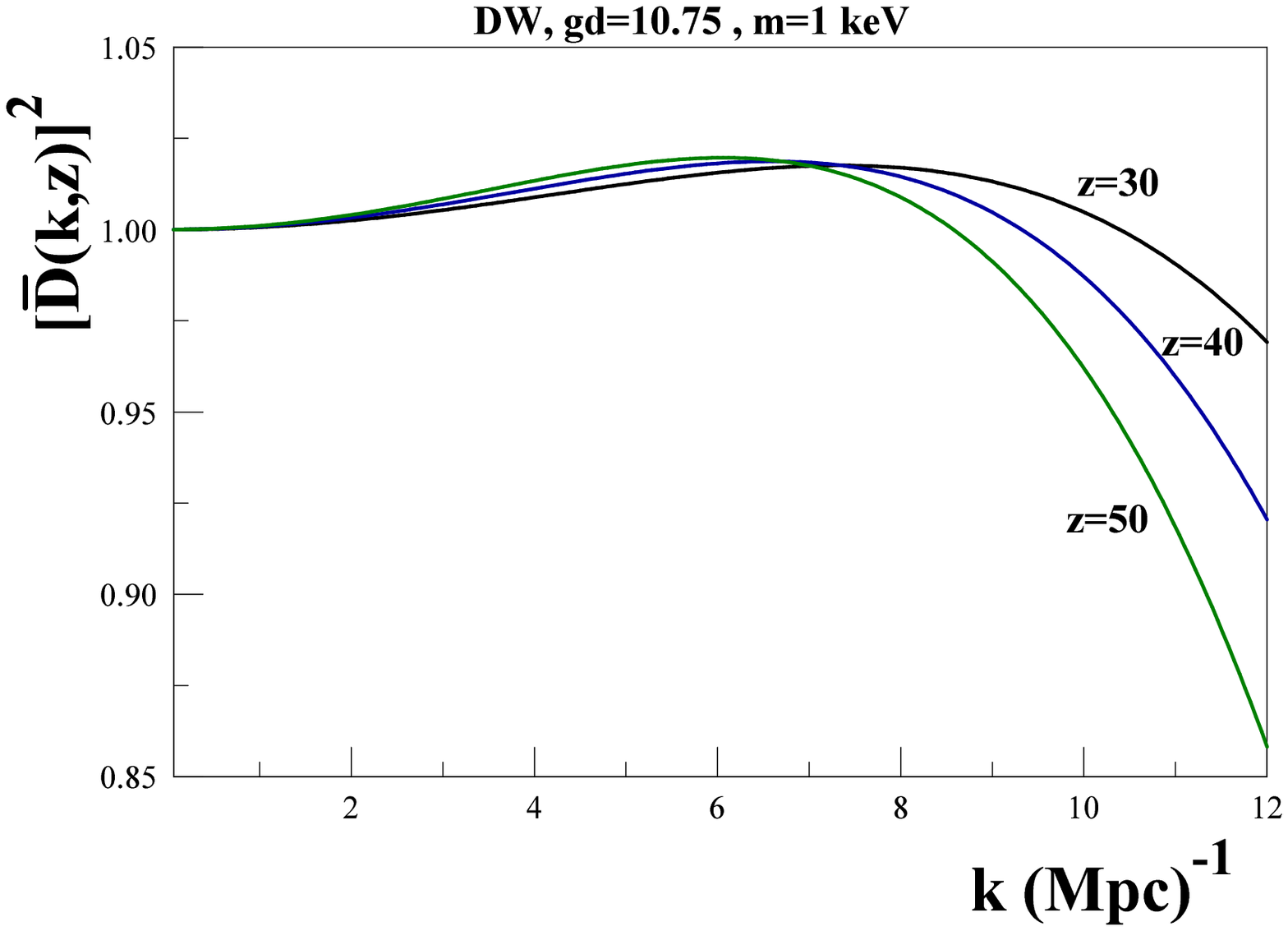}
\includegraphics[height=8cm,width=8cm,keepaspectratio=true]{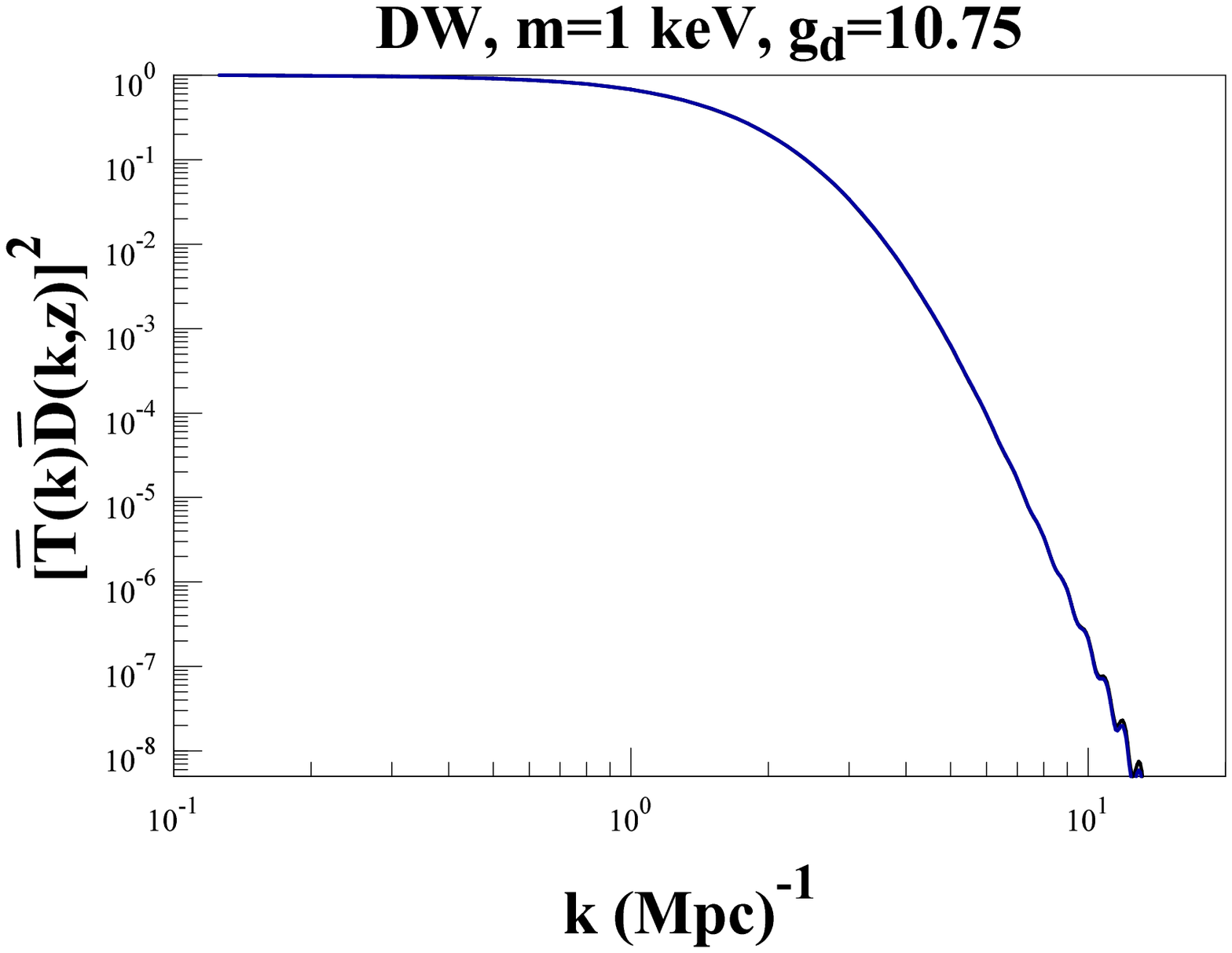}
\includegraphics[height=8cm,width=8cm,keepaspectratio=true]{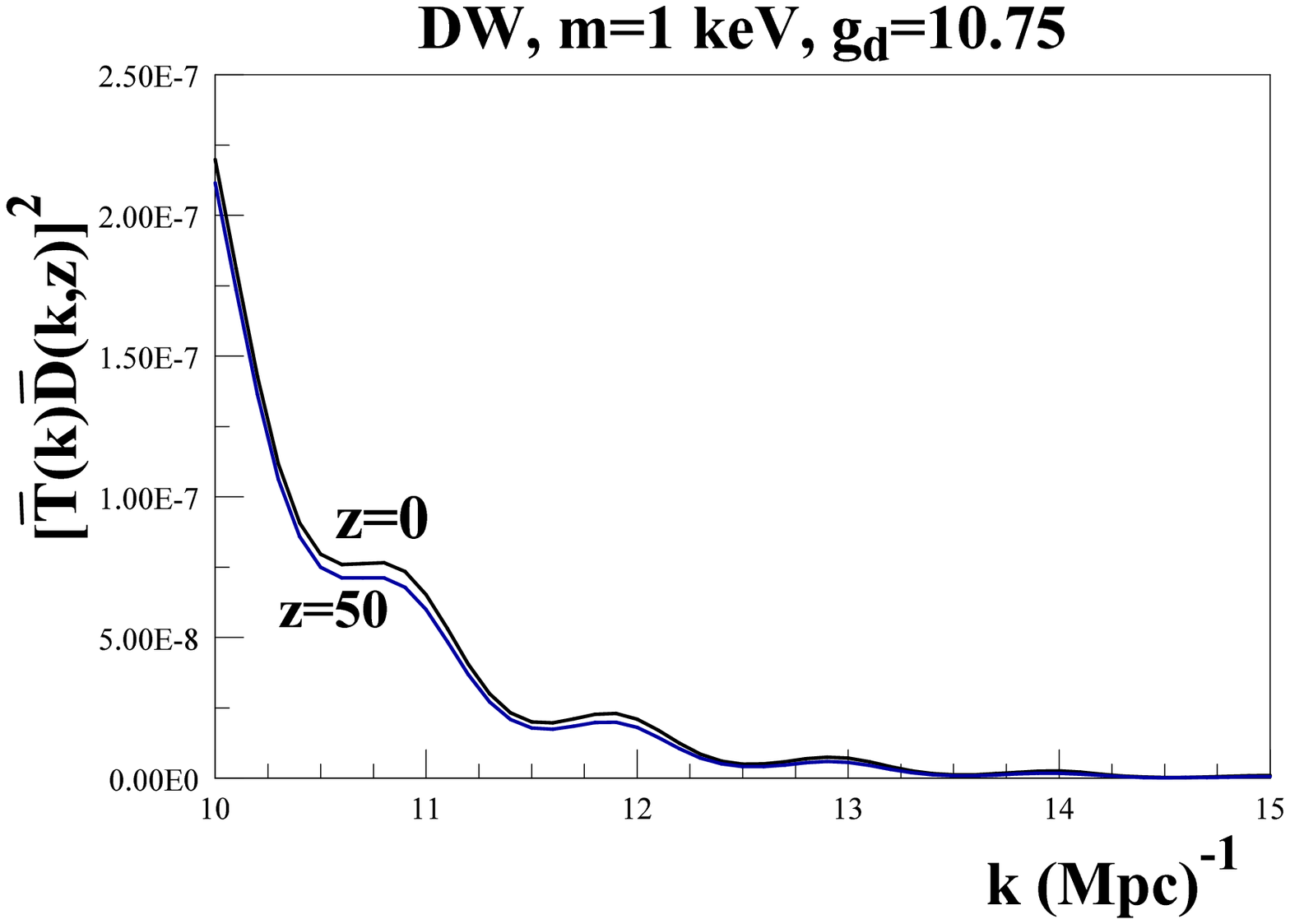}
\includegraphics[height=8cm,width=8cm,keepaspectratio=true]{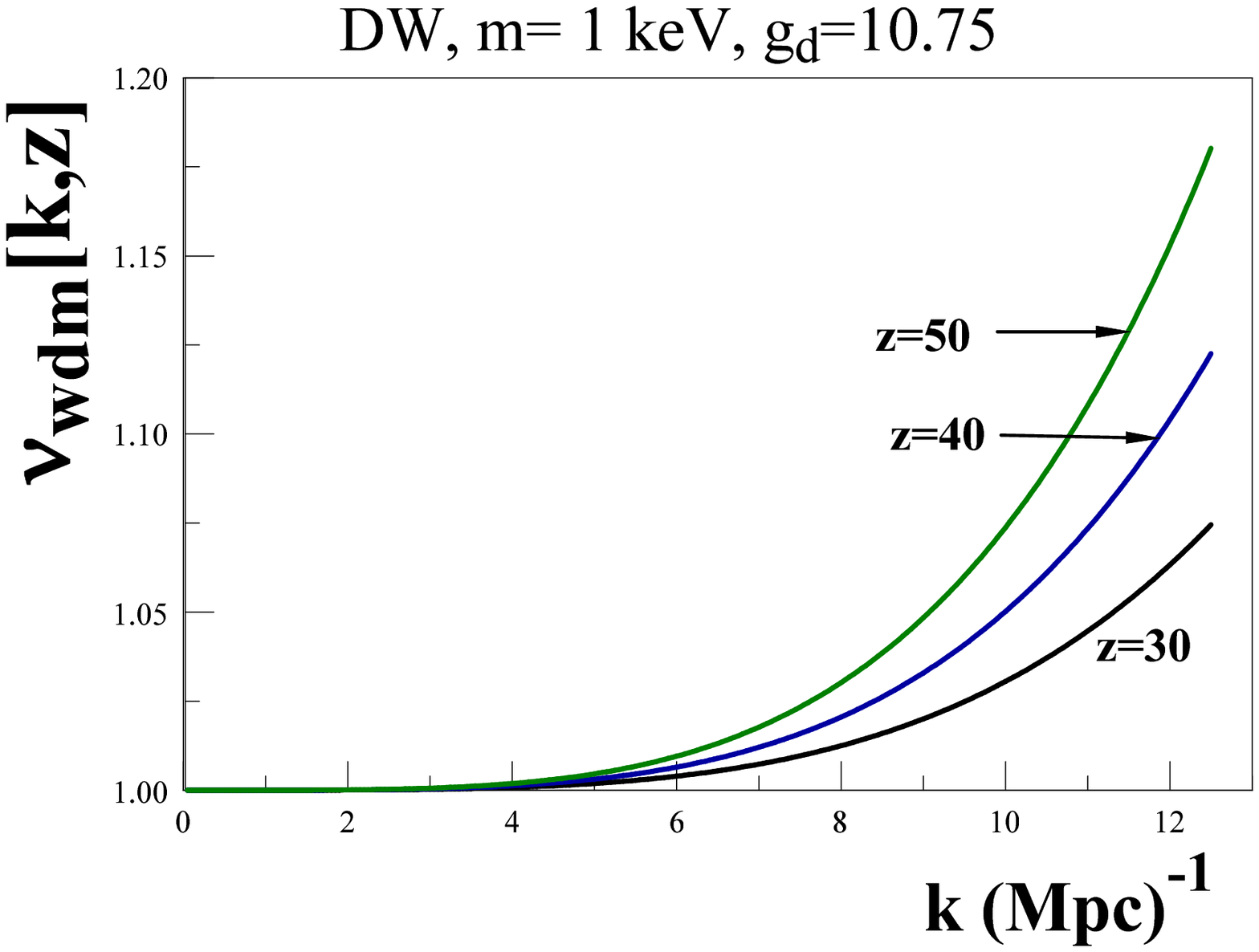}
\caption{DW, $m=1~\mathrm{keV}, g_d=10.75$: upper left panel
$\overline{D}[k,z]$ and
$[\overline{T}(k)\overline{D}(k,z)]^2$, lower panel:
small scale region of $[\overline{T}(k)\overline{D}(k,z)]^2$ and
  $\mathcal{V}_{wdm}[k,z]$ all for $z=30,40,50$.}   \label{fig:dw}
\end{center}
\end{figure}

   The (DW) case is displayed in fig.(\ref{fig:dw}): the fig. for $[ \overline{D}[k,z]]^2$ for the ``standard'' value $g_d=10.75$\cite{dw} clearly shows the crossover from an early enhancement to a later suppression of the power spectrum as a consequence of the contribution from peculiar velocity at small scales. For $m=1\,\mathrm{keV}$ (the value used in the figure)
  $k_{fs} = 5.44\,(\mathrm{Mpc})^{-1}$, and the figure clearly shows that the crossover from enhancement to suppression occurs at $k\approx 1-2\,k_{fs}$ for $30\leq z \leq 50$ . The corrections from $\overline{D}[k,z]$ are not resolved in the log-log scale, however a linear-linear display of the region $k  \gtrsim 2\,k_{fs}$ reveals the $10-15\,\%$ suppression of the power spectrum. This range of small scales is where the power spectrum develops the oscillatory behavior associated with the (WDM) acoustic oscillations discussed in ref.\cite{smallscale}.

\vspace{2mm}

\textbf{(BD) sterile neutrinos:}

\vspace{2mm}

 Sterile neutrinos produced by the decay of scalar or vector bosons
at the electroweak scale\cite{boysnudm,jun} are \emph{colder} for two reasons, i) their decoupling occurs when $g_d \sim 100$ and they do not reheat when the entropy from other degrees of freedom is given off to the thermal plasma, ii) their distribution function (\ref{fbd}) is more enhanced at small momentum thereby yielding smaller velocity dispersion. For this species

 \be \overline{y^2} = 8.509 ~~;~~ \beta_2 = 2.890  ~~;~~ k_{fs} = 14.107 \, \Big( \frac{m}{\mathrm{keV}}\Big)\,\Big( \frac{g_d}{100}\Big)^\frac{1}{3} \,\Big(\mathrm{Mpc} \Big)^{-1}\label{overysbd}\ee

\begin{figure}[ht]
\begin{center}
\includegraphics[height=8cm,width=8cm,keepaspectratio=true]{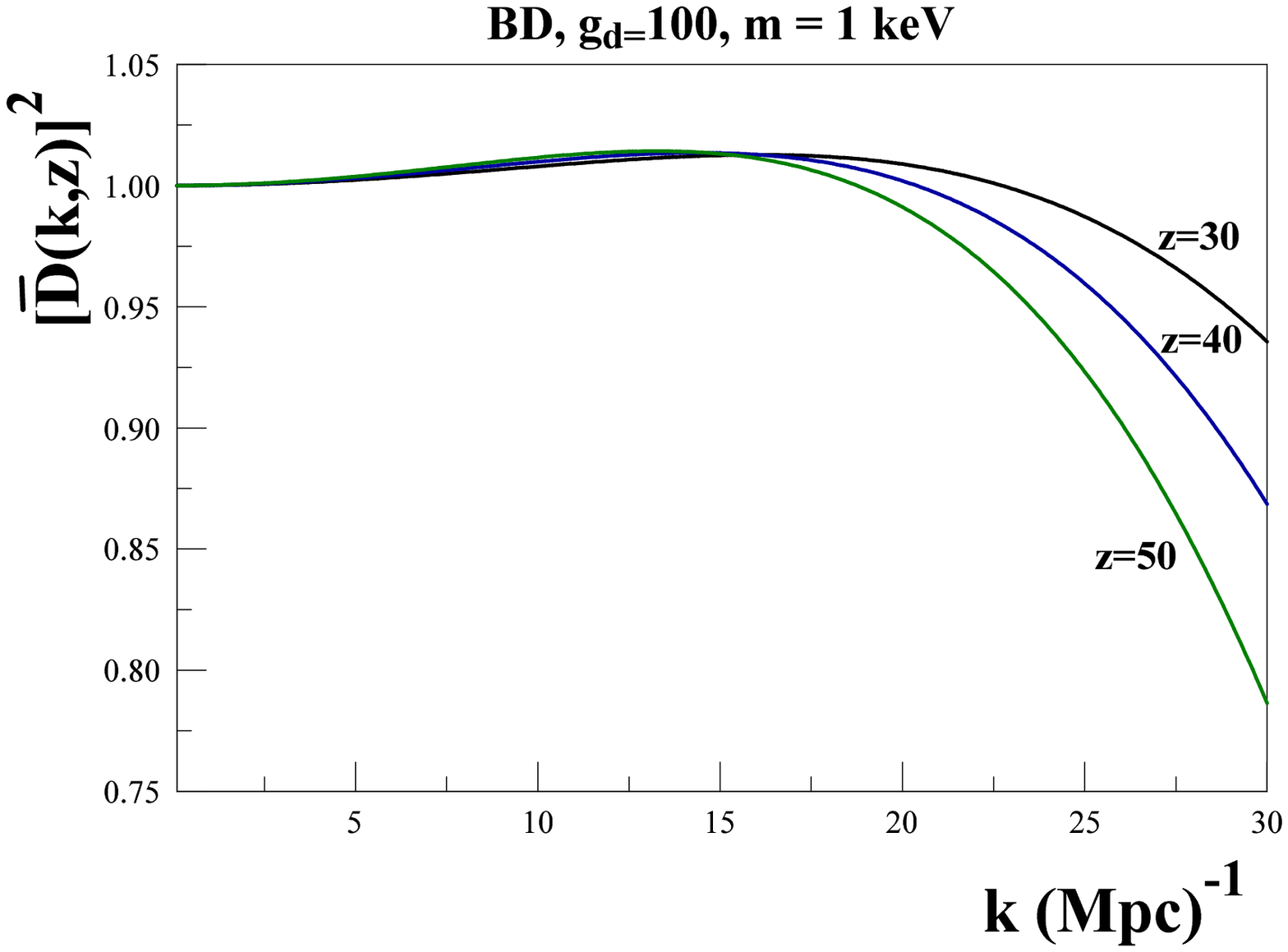}
\includegraphics[height=8cm,width=8cm,keepaspectratio=true]{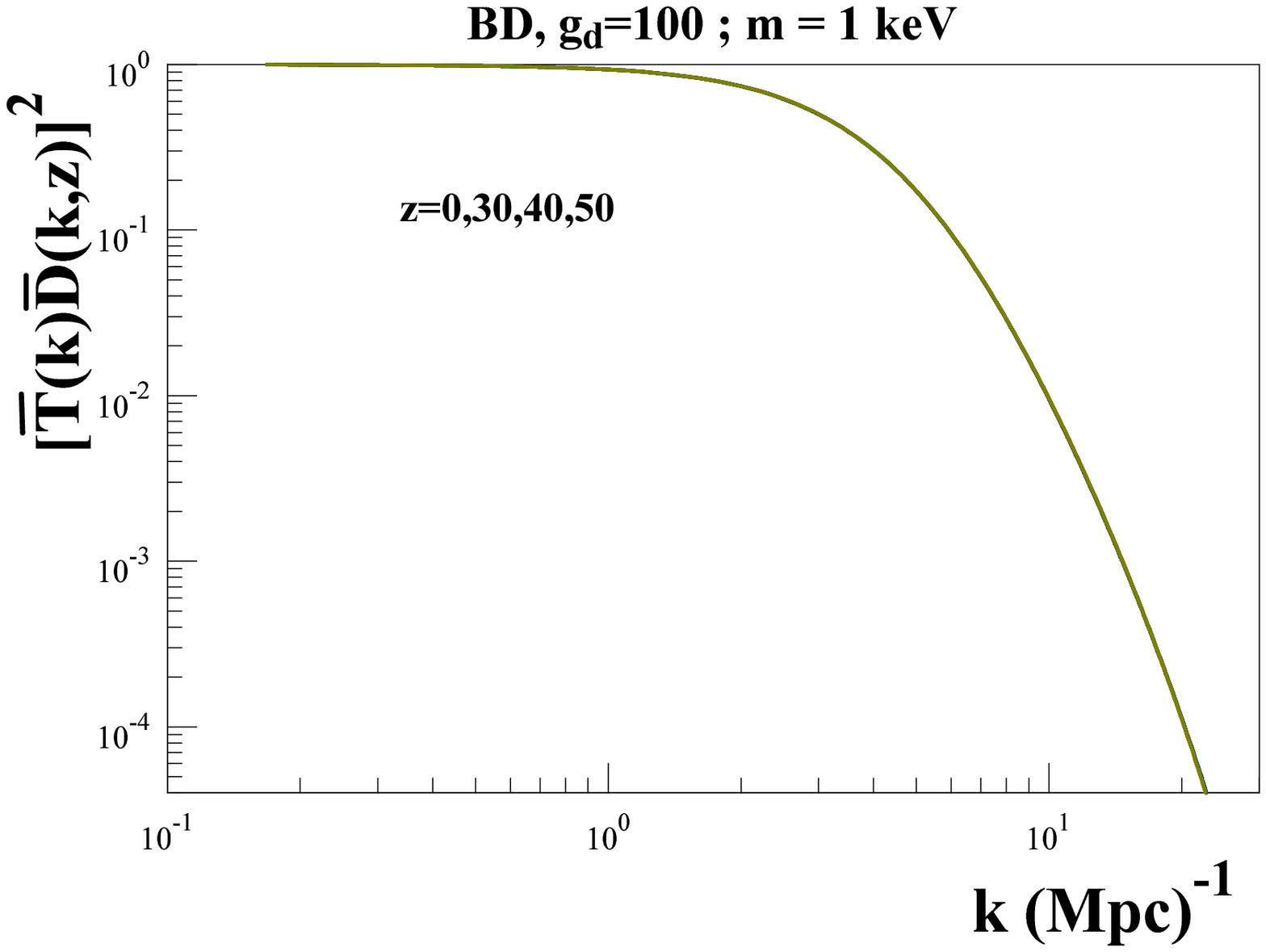}
\includegraphics[height=8cm,width=8cm,keepaspectratio=true]{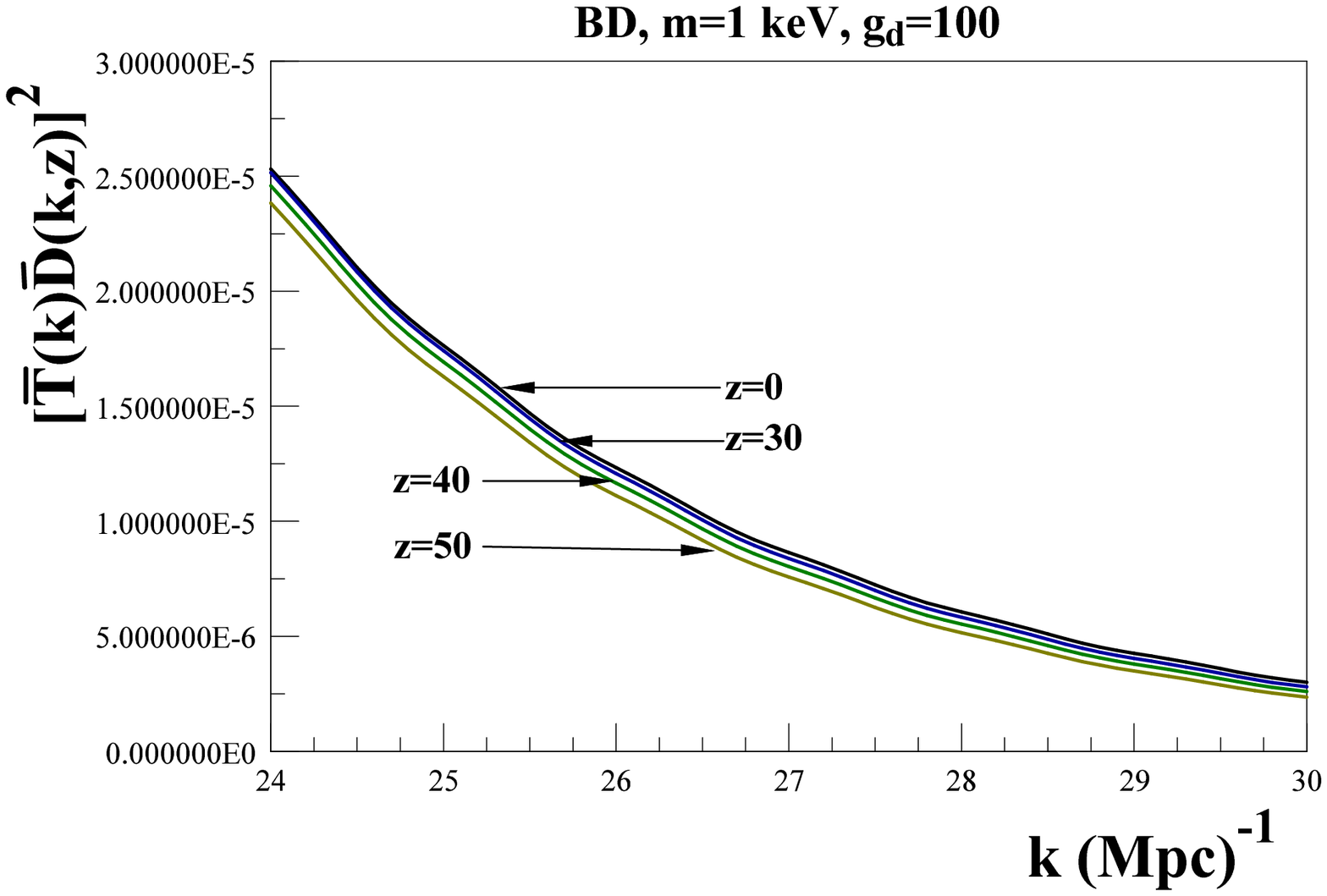}
\includegraphics[height=8cm,width=8cm,keepaspectratio=true]{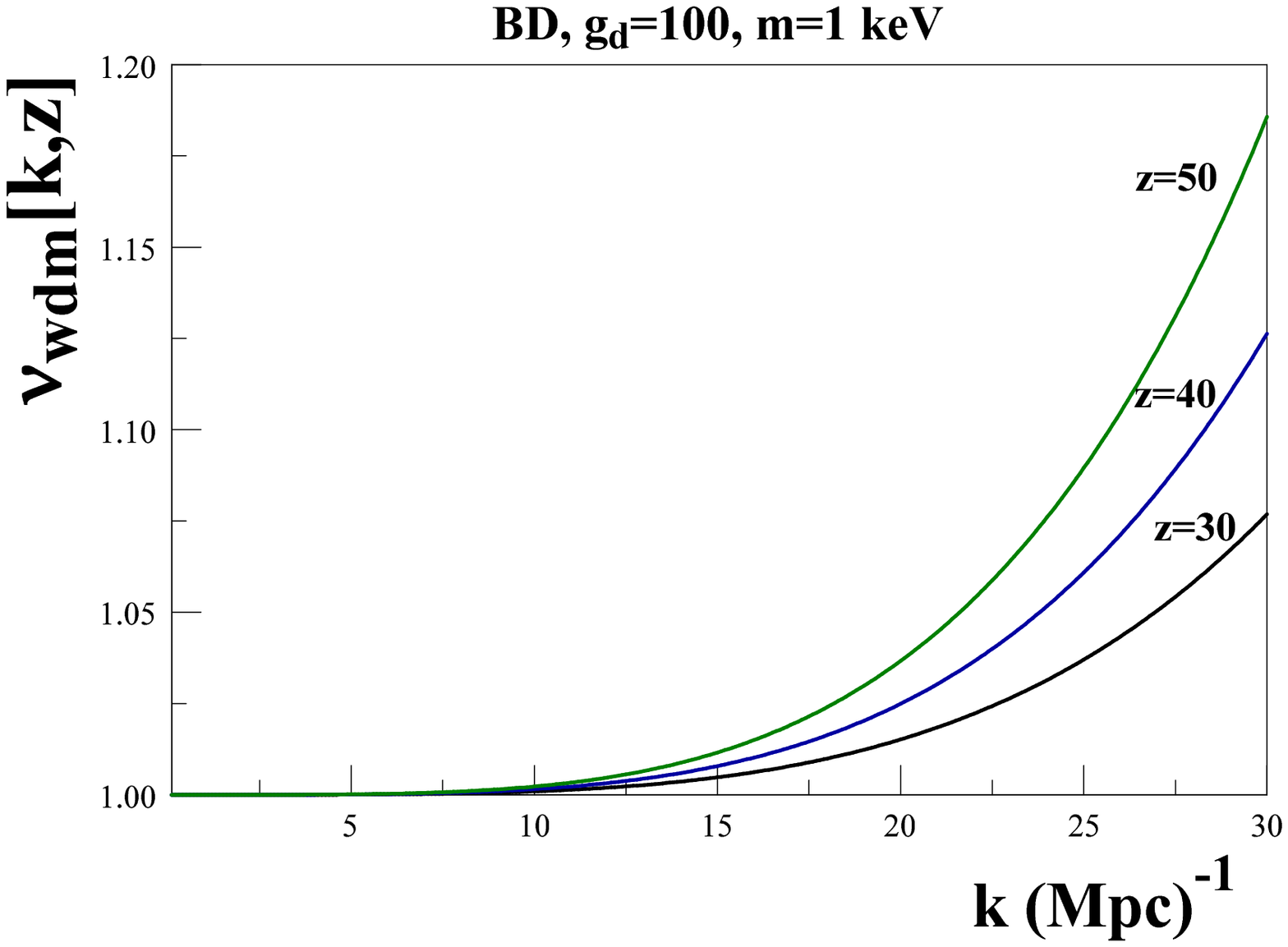}
\caption{BD, $m=1~\mathrm{keV}, g_d=100$: upper left panel
$\overline{D}[k,z]$ and
$[\overline{T}(k)\overline{D}(k,z)]^2$, lower panel:
small scale region of $[\overline{T}(k)\overline{D}(k,z)]^2$ and
  $\mathcal{V}_{wdm}[k,z]$ all for $z=30,40,50$.}   \label{fig:bd}
\end{center}
\end{figure}

 This case is displayed in fig.(\ref{fig:bd}): the fig. for $[ \overline{D}[k,z]]^2$ for  $g_d=100$ (corresponding to
 freeze-out at the electroweak scale) also shows the crossover from an early enhancement as a
 consequence of free streaming to a later suppression of the power spectrum as a consequence of the extra
 contribution  to free streaming from peculiar velocity at small scales. For $m=1\,\mathrm{keV}$ (the value used in the figure)
  $k_{fs} = 14.107\,(\mathrm{Mpc})^{-1}$, and the figure clearly shows that, again,
  the crossover from enhancement to suppression occurs at $k\approx 1-2\,k_{fs}$ for $30\leq z \leq 50$ .
  The corrections from $\overline{D}[k,z]$ are not resolved in the log-log scale of the power spectrum,
   however a linear-linear display of the region $k  \gtrsim 2\,k_{fs}$ reveals the $10-15\,\%$ suppression of the power spectrum. In this region the figure displays a hint of the (WDM) acoustic oscillations discussed in ref.\cite{smallscale}. As discussed in ref.\cite{smallscale} the smaller amplitude of the (WDM) acoustic oscillations as compared to the (DW) case are a reflection of the fact that (BD) sterile neutrinos are \emph{colder} as explained above.

In both these cases, we see that there is a suppression of the power
spectrum for $z\sim 30-50$ in the small scale region $k \simeq
(1-2)\,k_{fs}$ and an enhancement of the peculiar velocity in the
same region, both effects are at the $10-15\%$ level and clearly
correlated: the larger peculiar velocity \emph{adds} to free
streaming depressing the power spectrum. Although these effects are
at the level of few percent, it is conceivable that they may be
magnified by the inherent non-linearities in the process of
gravitational collapse, perhaps leading to important consequences
for galaxy formation in N-body simulations.

\section{Conclusions}

Motivated by recent and forthcoming N-body simulations of galaxy formation in (WDM) scenarios, we set out to study the redshift corrections to the matter and  peculiar velocity power spectra  and corrections to the  phase space density from gravitational perturbations  in the region $30 \leq z \leq 50$. This is the region in redshift where N-body simulations set up initial conditions and the dark energy component can be safely neglected.

Drawing from results in ref.\cite{smallscale}, we implemented a perturbative expansion for the redshift and
scale dependence of the distribution function, matter density perturbations and coarse grained phase space density valid for $z/z_{eq} \ll 1$ and a wide range of scales, up to leading logarithmic order in the scale factor.

We find that for (WDM) the redshift dependence is determined by $\beta_2$, the \emph{kurtosis} of the unperturbed distribution function after freeze-out, with an \emph{enhancement} of  of the peculiar velocity power spectrum and autocorrelation function at larger redshift for $\beta_2>1$. This enhancement in the peculiar velocity hastens free streaming and leads to a further suppression of the matter power spectrum for $k > (1-2)\,k_{fs}$, where $k_{fs}$ is the free streaming wavevector.  For (WDM) gravitational perturbations lead to a \emph{suppression} of the statistical fluctuations of velocities when   $\beta_2 > 5/3$.

We also study the linear corrections to  the coarse grained phase space density
introduced in refs.\cite{dalcanton1,hogan,coldmatter,darkmatter,hectornorma,boysnudm}
 resulting from gravitational perturbations. We find that whereas these vanish for (CDM)
  resulting in a constant (coarse grained) phase space density, (WDM) perturbations lead
  to a logarithmic growth with scale factor as a consequence of the suppression of statistical fluctuations if
  $\beta_2 > 1.19$.

Two specific examples of (WDM) candidates are studied in detail: sterile neutrinos produced
non-resonantly either via the Dodelson-Widrow mechanism\cite{dw} or via the decay of scalar or
vector bosons at the electroweak scale\cite{boysnudm,jun}. In these cases we find that the corrections
to the power spectra of matter and peculiar velocities are of order $10-15\,\%$ for scales
$k \simeq (1-2)\,k_{fs}$ and redshifts
$z \simeq 30-50$.

\vspace{2mm}

\textbf{Impact on the bounds on the mass:} The scale and redshift dependence of the power spectra
are encoded in the \emph{effective} matter and velocity power spectra $P_{\delta}(k)\overline{D}^{\,2}[k,z]~;~
P_{\delta}(k)\mathcal{V}[k,z]$ with $\overline{D}[k,z]~;~\mathcal{V}[k,z]$ given by eqns. (\ref{overDofz},\ref{bigTau}) respectively.

 To assess the impact of
the above results on the bounds on the mass of the (WDM) particle
consider two N-body simulations with a particle of the \emph{same
mass} both setting up initial conditions at the same $z \simeq
30-50$, one with the matter and peculiar velocity power spectra at
$z \sim 1$ and the other with the spectra corrected by the scale and
redshift dependent factors obtained above. If $\beta_2>1$   the corrected matter
power spectrum features the (WDM) \emph{suppression } and the
peculiar velocity power spectrum features the (WDM)
\emph{enhancement} for $k \gtrsim k_{fs}$ found above. These effects at small scales
are akin to the suppression of density fluctuations and enhancement of velocity dispersion associated with a  \emph{lighter} particle for the un-corrected
power spectra. This is because a lighter particle features a smaller
$k_{fs}$ and a larger velocity dispersion. Therefore these
corrections allow \emph{larger masses} to describe the \emph{same}
large scale structure output from the N-body simulations as compared
to the un-corrected power spectra.  Thus one aspect of the
corrections is to allow larger mass (WDM) particles, thereby
relaxing the bound on the mass, at least for those models for which
$\beta_2 >1$. However, this is not all there is to the corrections, because the
coarse-grained phase space density \emph{increases}, which would correspond to a colder particle with
smaller velocity dispersion. Thus the net effect of the corrections cannot be simply characterized as being
described by an increase or decrease of the mass of the particle and ultimately must be understood via
a full N-body simulation.

Although these corrections are relatively small,  non-linearities
arising from gravitational collapse \emph{may} result in a
substantial amplification of these effects, if this is the case, and
only large scale N-body simulations with the corrected power spectra
can assess this possibility, then it is conceivable (and expected)
that the bounds on the mass of the (WDM)
 particle may need substantial revision.

The results obtained here suggest a breakdown of perturbation theory
either at large redshift and or small scales $k \gg k_{fs}$, this
is clearly an artifact of the expansion, the integral in
(\ref{intdifeq2}) which yields the logarithmic contribution is
bounded and well behaved both in the small scale and $u \rightarrow
u_{NR}$ limits\cite{smallscale}. However a systematic study of
smaller scales and or larger redshifts would require a full
numerical solution of the integro-differential equation
(\ref{intdifeq2}). If future N-body simulations find that the
corrections obtained here do modify the dynamics of large scale
structure formation in (WDM) models substantially, such a study may
be worthy of consideration.

\acknowledgements The author is partially supported by NSF grant
award
 PHY-0852497.

\end{document}